# Unified Equations of Boson and Fermion at High Energy and Some Unifications in Particle Physics

Yi-Fang Chang

Department of Physics, Yunnan University, Kunming, 650091, China

(e-mail: yifangchang1030@hotmail.com)

**Abstract**: We suggest some possible approaches of the unified equations of boson and fermion, which correspond to the unified statistics at high energy. A. The spin terms of equations can be neglected. B. The mass terms of equations can be neglected. C. The known equations of formal unification change to the same. They can be combined each other. We derive the chaos solution of the nonlinear equation, in which the chaos point should be a unified scale. Moreover, various unifications in particle physics are discussed. It includes the unifications of interactions and the unified collision cross sections, which at high energy trend toward constant and rise as energy increases.

## 1. Introduction

Various unifications are all very important questions in particle physics. In 1930 Band discussed a new relativity unified field theory and wave mechanics [1,2]. Then Rojansky researched the possibility of a unified interpretation of electrons and protons [3]. By the extended Maxwell-Lorentz equations to five dimensions, Corben showed a simple unified field theory of gravitational and electromagnetic phenomena [4,5]. Hoffmann proposed the projective relativity, which is led to a formal unification of the gravitational and electromagnetic fields of the general relativity, and yields field equations unifying the gravitational and vector meson fields [6]. Swann discussed the advantage of unifying the concepts of rest energy and kinetic energy which gives the mass-energy relation a more clear-cut meaning [7]. In fact, Schrodinger equation is unified to describe various bosons and fermions. In the Heisenberg spinor unified field theory [8,9], if the unified equation has a rest mass, it will become to a nonlinear Dirac equation

$$\gamma_\mu \partial_\mu \psi + m\psi - l_0^2 \psi(\psi^+ \psi) = 0. \qquad (1)$$

Recently, Shrock constructed and studied gedanken models with electroweak-singlet chiral quark fields, which exhibit several distinctive properties, and could arise as low-energy limits of grand unified theories [10]. Mohanty, et al., showed natural inflation driven by pseudo-Nambu-Goldstone bosons in the grand unified theory scale [11]. Jentschura, et al., presented a unified treatment of anharmonic oscillators of arbitrary even and odd degree [12]. Based on a unified quantum field theory of spinors assumed to describe all matter fields and their interactions Kober constructed the space-time structure of general relativity according to a general connection within the corresponding spinor space [13]. Bazzocchi presented a grand unified model based on SO(10) with a $\Delta(27)$ family symmetry with an extended seesaw mechanism, in which fermion masses and mixings are fitted and agree well with experimental values [14]. Martin computed nonuniversal gaugino masses from F-terms in nonsinglet representations of SO(10) and E6 and their subgroups, extending well-known results for SU(5) [15]. Donoghue and Pais motivated and explored the possibility that extra SU(N) gauge groups may exist independently of the standard model groups, and studied the running of the coupling constants as potential evidence for a common origin of all the gauge theories [16]. Mimura, et al., studied three families of chiral fermions unified in higher dimensional supersymmetric gauge theory [17].

Based on the nonlinear Klein-Gordon equation and nonlinear Dirac equations, by using the method of the soliton-solution, we derived Bose-Einstein and Fermi-Dirac distributions, respectively, and both distributions may be unified by the nonlinear equation [18]. Moreover, we proposed a new type of soliton equation, whose solutions may describe some statistical distributions, for example, Cauchy distribution, normal distribution and student's t distribution, etc. The equation possesses two characters. Further, from an extension of this type of equation we may obtain the exponential distribution, and the Fermi-Dirac distribution in quantum statistics [18].

**2. Unified equations of boson and fermion at high energy**

At high energy Bose-Einstein (BE) and Fermi-Dirac (FD) statistics are unified [19], so the corresponding equations of boson and fermion should be unified at high energy [20]. We propose the three possible methods of the unified equations.

(A) Method. We assumed that the base of the unified statistics is that the spin differences of various particles may be neglected at high energy [19]. So the spin terms are very small and may be neglected in the equations, when they compare with other terms. It corresponds to that Pauli equation of non-relativity reverts to Schrodinger equation, which unified to describe various bosons and fermions.

When Dirac equation becomes a second order equation in electromagnetic field, it is [21]:

$$[(i\partial - eA)^2 - \frac{e}{2}\sigma_{\mu\nu}F^{\mu\nu} - m^2]\psi = 0. \tag{2}$$

At high energy the energy of spin interaction is very small,

$$H' = -(\frac{e\hbar}{2mc}\sigma)B << H_0, \tag{3}$$

$H'$ can be neglected. So the equation is unified to Klein-Gordon equation. In an external color field the QCD equation of a relativistic colored spinning particle is [22]:

$$[(i\hbar\partial^\mu + \frac{g}{c}A^\mu)^2 - (\frac{g\hbar}{2c})\sigma^{\mu\nu}F_{\mu\nu} - m^2c^2]\psi = 0. \tag{4}$$

When the energy $S^{\mu\nu}F_{\mu\nu}$ of spin interaction is neglected, Eq.(4) returns to Klein-Gordon equation, too.

In (A) method, boson and fermion and their equations trend toward unification.

(B) Method. At high energy the rest mass is lesser in $E^2 = c^2p^2 + m_0^2c^4$, if

$$m_0c^2 << E, \tag{5}$$

so $m_0c^2$ can be neglected, then $E^2 \approx c^2p^2$. The equations of particles whose spins are 1/2, and 0, 1 return to

$$\gamma_\mu \partial_\mu \psi \approx 0, \tag{6}$$

$$\Box \varphi \approx 0, \quad \Box A_\mu \approx 0. \tag{7}$$

They are Weyl neutrino equation, the photon-like (spin-0) equation and Maxwell equation. The spins of neutrino and photon, etc., are different, they are first and second order differential equations, respectively, in the present theory. Although according to the equivalence between $E^2 = c^2 p^2$ and $E = \pm cp$, neutrino and photon, etc., should be the same. The nonlinear equation (1) turns out Heisenberg spinor unified field equation.

In (B) method, at high energy boson and fermion are analogous with Goldstone (spin-0 and massless), photon and neutrino, and possess some characters of these particles.

Based on the inequalities (3) and (5), the scaling value of unification may be evaluated.

(C) Method. The known equations of formal unification [23,24] change to the same, for example, Dirac-Fiezz-Pauli equations, Bargmann-Wigner equations and Rarita-Schwinger equations. Especially, for Kemmer equations

$$(\beta_\mu \partial_\mu + m)\psi = 0, \tag{8}$$

and Dirac equations

$$(\gamma_\mu \partial_\mu + m)\psi = 0, \tag{9}$$

in which the matrix $\beta_\mu$ and $\gamma_\mu$ will perhaps be the same at high energy.

Three methods can be combined each other. For instance, if (C)+(B), so Bergmann-Wigner equations will trend toward massless. Such at high energy the spin of particle will be able only to point in the direction of motion [24]. If (A)+(B), so all of particles will trend toward neutrino, Goldstone particle, and photon, further, these massless particles will trend yet toward the same at high energy. Reversely, the photon at high energy is analogous with hadron.

Let the probability density $\rho = \psi^+ \psi = \overline{\psi} \gamma_4 \psi$, Eq.(1) becomes an ordinary differential equation [25]:

$$\frac{d\rho}{d\eta} = l_0^2 \rho(2\rho - 1), \tag{10}$$

It corresponds to a difference equation [25]

$$X_{n+1} = 1 - \frac{1}{4} l_0^4 X_n^2. \tag{11}$$

Further, Eq.(1) may be extended to [18]:

$$\gamma_\mu \partial_\mu \psi + m\psi = nb\psi(\psi^+ \psi), \tag{12}$$

It corresponds to a similar difference equation

$$X_{n+1} = 1 - \mu X_n^2. \tag{13}$$

Eq.(13) has always a chaos solution, as long as $n \neq 0$. So the unification of BE and FD

distributions is not only formal but also deeper meaning [19,18]. When n=1 and -1, only the X direction is reversed, but both topological properties are the same. The parameter $\mu$ is independent of n. Therefore, various values $\mu_1$, $\mu_2$,... $\mu_\infty$ of the branches are still the same for different n. It shows that the chaos theory is universal and unified for boson and fermion in many aspects. Both particles trend toward chaos, and reach to unification under the same condition $\mu_\infty =1.40115...$ This is not only a type of supersymmetry between boson and fermion, but also is agreement with the experiments of the multiplicity and its distributions, etc., which are independent of the types of particles. We think that the chaos theory may just describe the multiparticle production, and when energy reaches to up a certain threshold, chaos will appear [20,26]. The chaos point $\mu_\infty$ should be a unified scale.

At high energy there is the asymptotic free in the quark-parton model [27,28], such quarks are free, or weakness-interaction. So at high energy the collision of particles can revert to quark-quark interactions. The corresponding equations are:

$$(\gamma_\mu \partial_\mu + m)\psi = J . \tag{14}$$

Here J represents interactions, which correspond to the potential. Using the method of the soliton-solution, let $\xi = \dfrac{\gamma_\alpha x_\alpha - \gamma_0 ut}{1+u}$, Eqs.(14) become:

$$\frac{d\psi}{d\xi} + m\psi = J . \tag{15}$$

In order to simplify Eq.(15), assume that J is not the function of $\psi$, so Eq.(15) is just a first order linear ordinary differential equation. In the momentum representation, if we apply the (B) method $m \approx 0$, and J is independent of $p_i$, so

$$\psi = J(Ap + BE) + C_0 . \tag{16}$$

The method of quantum field theory is combined with the solution of equation in quantum mechanics, and then we obtained a cross section of collision [20]:

$$\sigma = C |\psi|^2 . \tag{17}$$

1. If let $J \approx 0$ (first approximation), which corresponds to the asymptotic free, so $\psi \approx C_0$. It is a particular solution of the equation of (B) method at high energy, therefore,

$$\sigma = \text{constant}. \tag{18}$$

2. Although J is small, but energy and momentum are very large, so

$$\sigma = C[J(Ap + BE) + C_0]^2 . \tag{19}$$

The cross sections rise as p and E increase at ultrahigh energy. Using a simple and approximate way, we may explain the experimental results: The collision cross sections of various particles at high energy have trended toward constant and toward slow rise at ultrahigh energy.

**3. Some unifications in particle physics**

We summarized that many cross sections of collisions at high energy may be unified, and be described by the $\Gamma$ distribution [29-31,19]:

$$y = \frac{\beta^\alpha}{\Gamma(\alpha)} x^{\alpha-1} \exp(-\beta x). \qquad (20)$$

Acosta, et al., analysed soft and hard interactions in pp collisions at $\sqrt{S}$ = 1800 and 630GeV, in which the figure of the scaling variable $Z = N*_{ch}/<N*_{ch}> - <N*_{ch}> P(Z)$ obeys the $\Gamma$ distribution (20) [32].

At high energy the unification should appear in the jet and the multiple production [20,26] and so on. Indeed, the KNO scaling and the Dao scaling, etc., are the same for all of particles.

The equation of the $\Gamma$ distribution corresponds to the Pearson equation [19]:

$$\frac{dy}{dx} + \beta y = \frac{\alpha-1}{x} y, \qquad (21)$$

and

$$x^2 \frac{d^2 y}{dx^2} - (\alpha-1)x\frac{dy}{dx} + (\alpha-1)(\beta+1)xy - \beta^2 x^2 y = 0. \qquad (22)$$

Let $x=\ln S$, the $\Gamma$ distribution (20) becomes to

$$\sigma = A(\ln S)^{\alpha-1} S^{-\beta}. \qquad (23)$$

It included some known formulas on cross sections of collisions.

A simple nonlinear equation is:

$$\frac{dy}{dx} = y^q. \qquad (24)$$

This may extend to another equation

$$\frac{dy}{dx} = \sum_n y^n. \qquad (25)$$

It includes two simplified forms

$$\frac{dy}{dx} = a(y \pm y^2). \qquad (26)$$

Their solutions are

$$y = \frac{1}{e^{-ax+c} \mp 1}, \qquad (27)$$

which are Bose-Einstein distribution and Fermi-Dirac distributions, respectively.

Further, Eq.(24) may be developed to a form

$$\frac{dy}{dx} = [\frac{\alpha-1}{x} - \beta]y + \frac{n\beta}{a} x^{1-\alpha} y^2, \qquad (28)$$

which is also a generalized form of the Pearson equation (21). The solution of Eq.(28) is

$$y = \frac{ax^{\alpha-1}}{e^{\beta x+c} + n}. \tag{29}$$

This is the quantum Bose-Einstein and Fermi-Dirac distributions, respectively, for n=1 or n=-1. When n=0 or $e^{\beta x+c}$ >>n, it is the $\Gamma$ distribution (20), which is a statistics unifying the Bose-Einstein and Fermi-Dirac distributions for particles at high energy [19]. When the nonlinear term $\pm ay^2$ in Eq.(26) is neglected, a usual Maxwell-Boltzmann distribution is obtained.

It is known that many particles and resonances are unified by Regge trajectories [33], and mesons and baryons are unified by quarks, which are developed to preons or subquarks, even prepreons.

The unification of various interactions is always an important question in particle physics. Weinberg and Salam proposed a well-known electroweak theory unified the weak and electromagnetic interactions, whose unified gauge group is SU(2)$\otimes$U(1)=U(2) [34,35]. Further, various grand unified theories of the strong, weak and electromagnetic interactions are researched [36-42], whose pioneer is Bars-Halpern-Yoshimura model [36,37]. Pati and Salam proposed the unified lepton-hadron symmetry and a gauge theory $SU(2')\otimes U(1)\otimes SU(3'')$ of the basic interaction [38]. There is the same gauge group in Itoh-Minamikawa-Miura-Watanabe model [39]. A famous theory is the Georgi-Glashow SU(5) theory [40]. Moreover, there are Fritzsch-Minkowski SU(n)$\otimes$SU(n) (n=8,12,16) and SO(n) (n=10,14) unified interactions theories of leptons and hadrons [41], and a universal gauge theory model based on E(6) [42], and so on. Recently, Calmet, et al., showed grand unification and some enhanced quantum gravitational effects [43]. Blumenhagen investigated gauge coupling unification for F-theory SU(5) grand unified theories with gauge symmetry breaking via nontrivial hypercharge flux [44]. We proposed that the infinite gravitational collapse of any supermassive stars should pass through an energy scale of the grand unified theory. After nucleon-decays, the supermassive star will convert nearly all its mass into energy, and produce the radiation of grand unified theory. It may probably explain some ultrahigh energy puzzles in astrophysics, for example, quasars and gamma-ray bursts, etc. This is similar with a process of the Big Bang Universe with a time-reversal evolution in much smaller space scale and mass scale. In this process the star seems be a true white hole [45].

Einstein gravitational Lagrangian possesses two invariances: the GL(4,R) invariance of Einstein under coordinate transformations, and the SL(2,C) gauge invariant of Weyl. The strong interaction of quarks possesses internal SU(3) symmetry. From these symmetries Isham, Salam and Strathdee proposed the unified scheme on gravitational and strong interactions, whose gauge group is SU(3)$\otimes$SL(2,C)=SL(6,C) [46-48].

Based on the Weinberg-Salam unified electroweak theory and Isham-Salam-Strathdee strong-gravitational interactions unified scheme [46-48], in 1974 we proposed that the simplest unified gauge group of four-interactions must be GL(6,C) or its extension [49,20]. In 1977 Terazawa proposed also a unified gauge symmetry of GL(32N,C) or GL(12+2n,C) for combining the SU(16N) or SU(6+n) group of lepton-quark internal symmetry and the SL(2,C) Lorentz group of space-time symmetry (where N=1,2,3… and n=0,1,2,3…- ) [50]. Of course, this unified group is a non-compact group, and seems to construct a 'no-go' law. But, it is not only reasonable but also necessary, because the classical gravitational interaction corresponds to a non-compact group SL(2,C). Further, it should be that the gravitational theory is developed to the quantum gravity theory, for example, the supergravity and the loop quantum theory [51-53] and so on.

The supersymmetry theory describes a basic symmetry between bosons and fermions, and arouses the supergravity in the gravitational theory, and derives a superstring combining a string model. This is related with the unified theory of interactions. Recently, Barr and Raby proved minimal SO(10) unification in the supersymmetric grand unified theory [54]. Kakushadze and Tye researched the classification of three-family grand unification of SO(10), E(6), SU(5) and SU(6) models in string theory [55]. Das and Jain discussed dynamical gauge symmetry breaking in an $SU(3) \otimes U(1)$ extension of the standard model [56]. Albright and Barr discussed explicit SO(10) and $U(1) \otimes Z(2) \otimes Z(2)$ supersymmetric grand unified model for the Higgs and Yukawa sectors [57].

We researched some new representations of the supersymmetric transformations, and introduced the supermultiplets [58]. Then various formulations of bosons and fermions may be unified. On the one hand, the mathematical characteristic of particles is proposed: bosons correspond to real number, and fermions correspond to imaginary number, respectively. Such fermions of even (or odd) number form bosons (or fermions), which is just consistent with a relation between imaginary and real number. The imaginary number is only included in the equations, forms, and matrixes of fermions. It is connected with relativity. On the other hand, the unified forms of supersymmetry are also connected with the statistics unifying BE and FD statistics, and with the possible violation of Pauli exclusion principle; and a unified partition function is obtained [19,20]. A possible development is the higher dimensional complex space [58]. Triantaphyllou and Zoupanos searched strongly interacting fermions from a higher (4-12) dimensional $E(8) \otimes E(8)'$ unified gauge theory [59].

At high energy the both BE and FD statistics are unified [19], and the both equations are unified, so the characters of boson and fermion will trend toward the same, i.e., the both particles will be unified. This is consistent with the directions of the present theory, such as the unification of interactions and the supersymmetry and so on, and with the general tendency of experiments at high energy [20]. Furthermore, various unifications are connected each other. These unifications will determine the scaling values of unification more exactly, and will expect more phenomena, and will be applied to more regions. Any unification is symmetry and a simplified. It should be an important aspect of developed theories.